\documentclass[aps,prd,a4paper,twocolumn,showpacs,nofootinbib]{revtex4-1}

\usepackage{amssymb,amsmath,latexsym,mathrsfs}
\usepackage{graphicx,subfigure}
\usepackage{epsfig}
\usepackage{varioref,xr-hyper}
\usepackage{color}

\newcommand{\Mpc}{{\rm ~Mpc}}

\newcommand{\neff}{N_{\rm {eff}}}

\begin{document}

\title{The Fine Structure Constant and the CMB Damping Scale}

\author{Eloisa Menegoni$^{a}$}
\author{Maria Archidiacono$^{b}$}
\author{Erminia Calabrese$^{c}$}
\author{Silvia Galli$^{d}$}
\author{C. J. A. P. Martins$^{e}$}
\author{Alessandro Melchiorri$^{b}$}

\affiliation{$^a$ I.C.R.A. and INFN, Universit\`a di Roma ``La Sapienza'', Ple Aldo Moro 2, 00185, Rome, Italy}
\affiliation{$^b$ Physics Department and INFN, Universit\`a di Roma ``La Sapienza'', Ple Aldo Moro 2, 00185, Rome, Italy}
\affiliation{$^c$ Sub-department of Astrophysics, University of Oxford, Keble Road, Oxford, OX1 3RH, UK}
\affiliation{$^d$ Institut d`Astrophysique de Paris, UMR 7095-CNRS Paris, Universit\'e Pierre et Marie Curie, boulevard Arago 98bis, 75014, Paris, France}
\affiliation{$^e$ Centro de Astrof\'{\i}sica, Universidade do Porto, Rua das Estrelas, 4150-762 Porto, Portugal}

\begin{abstract}

The recent measurements of the Cosmic Microwave Background anisotropies at
arcminute angular scales performed by the  ACT and SPT experiments are probing
the damping regime of CMB fluctuations.  The analysis of these datasets unexpectedly
suggests that the effective number of relativistic degrees of freedom is larger than
the standard value of $\neff=3.04$, and inconsistent with it at more than two standard
deviations. In this paper we study the role of a mechanism that could affect the shape
of the CMB angular fluctuations at those scales, namely a change in the recombination
process through variations in the fine structure constant. We show that the new CMB
data significantly improve the previous constraints on variations of $\alpha$, with
$\alpha / \alpha_0 = 0.984 \pm 0.005$, i.e. hinting also to a more than two standard
deviation from the current, local, value $\alpha_0$. A significant degeneracy is present
between $\alpha$ and $\neff$, and when variations in the latter are allowed the
constraints on $\alpha$ are relaxed and again consistent with the standard value. 
Deviations of either parameter from their standard values would imply the presence
of new, currently unknown physics.
\end{abstract}
\maketitle

\section{Introduction}

The recent observations from Cosmic Microwave Background (CMB hereafter) satellite,
balloon-borne and ground based experiments (\cite{wmap7,act,acbar,spt}),
galaxy redshift surveys \cite{red} and luminosity distance measurements,
have fully confirmed the theoretical predictions  of the standard $\Lambda$CDM
cosmological model. This not only allows stringent constraints
on the parameters of the model but can be fruitfully used to constrain non
standard physics at the fundamental level, such as classes of elementary particle
models predicting a different radiation content in the Universe.

In this respect, an interesting discrepancy with the expectations of the standard
model has recently been uncovered in the small CMB scale measurements of the
ACT \cite{act} and SPT \cite{spt} experiments. Namely, the effective number of
relativistic degrees of freedom $\neff$  (see e.g. \cite{kolb} for a definition)
has been reported as higher (at more than two standard deviations) than the expected
standard value of  $\neff=3.046$ \footnote{This is the value expected
in the case of $3$ relativistic neutrinos species. The little deviation from
 $\neff=3$ takes into account effects from the non-instantaneous neutrino
decoupling from the primordial photon-baryon plasma (see e.g. \cite{mangano3046})}.
This result has been confirmed by several recent analyses of the ACT and
SPT datasets (see e.g. \cite{archi011,knox11,zahn,hamann11,giusarma11}).

While a confirmation from future measurements is clearly needed,
the current preference for $\neff \sim 4$ is stimulating a growing interest
since it could be explained in different ways, including an extra relativistic
particle at decoupling (as light axions or sterile neutrinos), extra dimensions, or
early dark energy. (see e.g. \cite{archi011} and references therein).

However it is important to stress that the current bounds on $\neff$ rely
on the assumption of a theoretical model. More recently the dependence
of the constraints on $\neff$ on the assumption of a flat universe
or a different dark energy component have been investigated by 
several authors (\cite{aaron,giusarma12,joudaki}). 
Here we revisit the issue, by obtaining analogous constraints
in the framework of a non-standard recombination process.

As expected (see e.g. \cite{knox11}) a variation of $\neff$ affects
the value of the Hubble parameter $H$ at recombination. This changes two very
important scales in CMB anisotropy physics: the size of the sound horizon and
the  damping scale at recombination. An approximate expression
for the damping scale is given by

\begin{equation} 
\label{damp}
r_d^2 = \left(2\pi\right)^2\int_0^{a_*} \frac{da}{a^3\sigma_T n_e
  H}\left[\frac{R^2 + \frac{16}{15}\left(1+R\right)}{6(1+R^2)}\right]
\end{equation}
where $n_e$ is the number density of free electrons, $\sigma_T$ is the Thompson
cross-section,  $a_*$ is the scale factor at recombination and $R=3\rho_b / (4\rho_\gamma)$
is proportional to the ratio between the baryon and photon densities.
It is clear that a change in $H$ could be compensated by a change
in $n_e$ and $a_*$ in order to keep the same damping scale. Consequently, a 
change in the recombination process, motivated by some non-standard and
unaccounted mechanism, could alter the current conclusions on $\neff$.

Possible changes in the recombination process have been
investigated by several authors. Dark matter annihilation, for example,
could significantly alter the evolution of the free electron density
$n_e$ by the injection of extra-ionizing photons around recombination
(see e.g. \cite{galli} and references therein). Another possible mechanism, which
we consider in this paper, is based on the hypothesis of a change
in the fundamental constants of nature, specifically the fine structure constant, $\alpha$.

Changing $\alpha$ modifies the strength of the electromagnetic interaction and therefore
modifies the formation of CMB anisotropies by changing the differential
optical depth  (i.e. the scattering rate) due to Thomson scattering between electrons and photons :

\begin{equation}
\dot{\tau} = x_{e}n_{e} c \sigma_{T} ,
\end{equation}

where  $\sigma_T$ is the Thomson cross section,  $x_{e}$ is the free electron fraction dependent on the temperature of the
electrons and therefore on the scale factor of the universe $a(t)$. The optical depth $\tau$ is then defined as the integral of the scattering rate over time.

These two combined processes change  the temperature at last scattering, $T^{*}$,
and $x_{e}(t_{0})$, the free electron fraction that remains after recombination, both of which influence the CMB anisotropies (see \cite{battye}).

The main imprint on CMB power spectrum is a shift in the modulation of the peak heights by baryon drag  determined by the relative density of baryons to photons at $\eta^{*}$, $R= 3\rho_b / (4\rho_\gamma) \sim T^{*^{ -1}}$.

A variation in $\alpha$ changes the photon diffusion damping length as well and the two effects combined lead to
subtle degeneracies between $\Delta \alpha/\alpha$ and $\neff$.

CMB anisotropies are therefore one of the canonical ways of constraining variations in
the fine structure constant in the early universe. They provide a measurement
of $\alpha$ at the epoch of recombination
(see e.g. \cite{avelino,Rocha,ichikawa,jap,petruta,menegoni,landau}), with a
current sensitivity the level of $\sim 1 \%$. In the most recent analysis, parametrizing
a variation in the fine structure constant as  $\alpha/\alpha_0$, where
$\alpha_0=1/137.03599907$ is the standard (local) value and $\alpha$ is the value during
the  recombination process, the authors of \cite{menegoni} used the five year WMAP data,
finding the constraint $0.987\pm0.012$ at $68 \%$ c.l.. Meanwhile, a recent analysis
of a large dataset of spectroscopic data from the VLT and Keck telescopes \cite{dipole}
is consistent with earlier claims of variations in the value of $\alpha$ at
parts-per-million level at redshifts $z\sim3$.

In view of this and the recent results from ACT and SPT is therefore extremely timely
to place new bounds on variations of $\alpha$ discussing also the possible
degeneracies with $\neff$. In this paper we indeed perform this kind of analysis,
including also possible variations in the abundance of primordial Helium
$Y_p$ that could similarly change the recombination process.
In the next section we describe the analysis method, in section III we present
our results while in Section IV we derive our conclusions.

\section{Analysis Method}

We perform a COSMOMC \cite{Lewis:2002ah} analysis combining the following CMB
datasets: WMAP7 \cite{wmap7}, ACBAR \cite{acbar}, ACT \cite{act}, and SPT \cite{spt},
and we analyze datasets out to $l_{\rm max}=3000$.  We also include
information on dark matter clustering from the galaxy power spectrum
extracted from the SDSS-DR7 luminous red galaxy sample
\cite{red}. Finally, we impose a prior on the Hubble parameter based
on the last Hubble Space Telescope observations \cite{hst}.

The analysis method we adopt is based on the publicly available Monte Carlo
Markov Chain package \texttt{cosmomc} \cite{Lewis:2002ah} with a convergence
diagnostic done through the Gelman and Rubin statistic.

We sample the following six-dimensional standard set of cosmological parameters,
adopting flat priors on them: the baryon and cold dark matter densities
$\Omega_{\rm b}$ and $\Omega_{\rm c}$, the Hubble constant $H_0$, 
the optical depth to reionization $\tau$,
the scalar spectral index $n_S$, and the overall normalization of the
spectrum $A_S$ at $k=0.002\Mpc^{-1}$. We consider purely adiabatic initial
conditions and we impose spatial flatness.
As discussed in the introduction we allow for variations in the fine
structure constant $\alpha/\alpha_0$ where $\alpha_0$ is the current,
local, value by modifying the RECFAST recombination subroutine
following the procedure described in \cite{menegoni}.

We also allow for variations in the effective number of relativistic degrees
of freedom $\neff$ and the primordial Helium abundance $Y_p$, otherwise fixed
at the values $\neff=3.046$ and $Y_p=0.24$, respectively. Since we are varying
also the Helium abundance, we considered variations in the fine structure constant
also in the process of Helium recombination. A $\sim 5\%$ change of $\alpha$
for Helium recombination changes the CMB angular spectra by less than
$0.5 \%$ up to $\ell =1500$. During reionization the fine structure constant
is fixed to the local standard value $\alpha=\alpha_0$.

We account for foregrounds contributions including three extra amplitudes:
the SZ amplitude $A_{SZ}$, the amplitude of clustered point sources $A_C$,
and the amplitude of Poisson distributed point sources $A_P$. 
We marginalize the contribution from point sources only for the ACT and SPT data,
based on the templates provided by \cite{spt}.  We quote only one joint amplitude
parameter for each component (clustered and Poisson distributed). The SZ amplitude
is obtained fitting the WMAP data with the WMAP own template, while for SPT and ACT
it is calculated using the \cite{Trac:2010sp} SZ template at 148 GHz; this differs 
from the analysis performed in \cite{spt} where no SZ contribution was considered for the WMAP data.

\section{Results}

As stated in the previous section, we perform three different analyses
always considering the same set of data but different number of parameters.
To the $6$ standard $\Lambda$-CDM parameters we cumulatively add as additional
free parameters the fine structure constant (first case), the
number of relativistic degrees of freedom $\neff$ (second case) and
the primordial Helium abundance $Y_p$ (third case). In Table \ref{standard} we
report the constraints on the cosmological parameters for these three scenarios.

\begin{table}[h!]
 \begin{center}
 \begin{tabular}{|l|c|c|c|}
 \hline
 \hline   
$ Parameter$  &  $\alpha/\alpha_0$  &  $ \alpha/\alpha_0$+$\neff$ & $\alpha/\alpha_{0}$+$\neff$+$Y_p$\\ 
\hline
$ \Omega_b h^2$ & $0.0218 \pm 0.0004 $ & $  0.0224 \pm 0.0005$ & $   0.0223 \pm   0.0007             $\\
$\Omega_{c} h^2$ & $0.1144 \pm 0.0034 $ & $0.1302 \pm  0.0095$  & $0.1303 \pm 0.0094$\\
$ \tau$ & $0.086 \pm 0.014$ & $ 0.088   \pm   0.015$ & $ 0.088    \pm    0.016                $\\
$ H_0$ & $68.9 \pm 1.4$ & $71.52  \pm 2.0$  & $71.8\pm2.1$\\
$\alpha/\alpha_0$ & $0.984 \pm 0.005$  & $0.990\pm0.006$  & $0.987\pm0.014$\\
$n_s$ & $0.976\pm0.013$  & $0.991\pm0.015$  & $0.992\pm0.016$\\  
$log[10^{10} A_s]$ & $3.193 \pm 0.037$  & $3.169\pm0.040$  & $3.167\pm0.042$\\
$A_{SZ} $ &  $< 2.00$ & $< 2.00$ & $< 2.00$\\ 
$A_C$  & $< 16.0$  & $<15.8$ & $< 14.8$\\  
$ A_P$  &     $< 24.7$ & $< 24.9$ & $< 22.4$\\
$\Omega_\Lambda$  & $0.7137  \pm 0.0070$ & $0.7020 \pm 0.0094$ & $0.704 \pm 0.013$\\ 
$Age/Gyr$ &   $13.76 \pm 0.24$ & $13.18 \pm 0.38$ & $13.15 \pm 0.37$\\ 
$ \Omega_m$  &     $0.2863 \pm 0.0070$ & $0.2980 \pm 0.0094$ & $0.296 \pm 0.013$\\ 
$ \sigma_8$ &   $0.836 \pm 0.023$ & $0.862 \pm 0.028$ & $0.859 \pm 0.034$\\ 
$z_{re}$  & $10.7 \pm 1.2$ & $11.0 \pm 1.3$ & $11.0 \pm 1.3$\\ 
$\neff$  & $ -  $ & $4.10_{-0.29}^{+0.24}$ & $4.19_{-0.35}^{+0.31}$\\
 $Y_p$ & $ -   $ & $  -  $  & $0.215 \pm 0.096$\\
 \hline
 $\chi^2_{min}$ & $7600.2 $ & $ 7596.8  $  & $ 7596.5    $\\    
 \hline
 \hline
 \end{tabular}
 \caption{MCMC estimation of the cosmological parameters 
 from the dataset described in the text. Results for the
 three analyses described in the text are reported.
 Upper bounds at $95 \%$ c.l. are reported for foregrounds
 parameters.}
 \label{standard}
 \end{center}
 \end{table}

\begin{figure}[h!]
\includegraphics[scale=0.4]{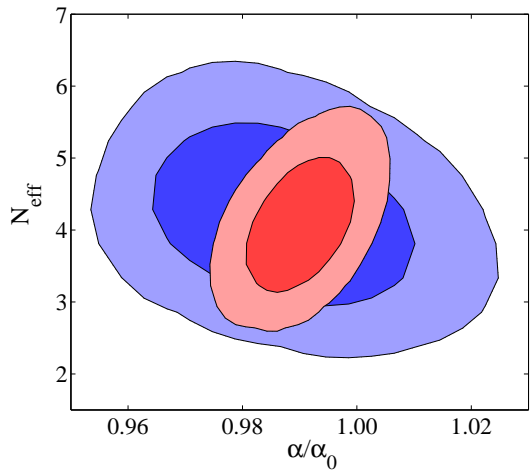}
\caption{Likelihood contour plot for $\alpha/\alpha_0$ vs $\neff$ at $68\%$ c.l. and $95\%$ c.l.
in the case of $Y_p=0.24$ (red smaller contours) and $Y_p$ allowed to vary (blue larger contours).}
\label{degenerazioni}
\end{figure}

As we can see the dataset considered prefers a value of
$\alpha/\alpha_0$ smaller than unity at more than two standard deviations
when both the $\neff$ and $Y_p$ are kept fixed at their standard values.
This result, while interesting, is to be expected since is clearly
driven by data preference for larger values of $\neff$.
Allowing for variations in $\neff$ significantly shifts the best fit
value for $\alpha / \alpha_0$, which is now consistent with the standard value. 
However, even in this case the best fit value for $\neff$ is still $\sim 4$, i.e.
allowing for variation in the fine structure constant enlarges the error bars on
$\neff$ of about $\sim 30\%$ but does not shift the best fit value towards the
standard result. The largest effect on $\alpha$ comes however when also the helium
abundance $Y_p$ is let to vary. In this case, indeed, the errors on
$\alpha$ are almost doubled.

We can better understand the impact of $Y_p$ on the determination
of $\alpha/\alpha_0$ by looking at Figure 1, where we plot the $2$-D likelihood
contours in the $\alpha/\alpha_0$-$\neff$ plane in the cases of $Y_p=0.24$ and
free $Y_p$. As we can see when the helium abundance is
fixed there is a clear but moderate degeneracy between
$\alpha/\alpha_0$ and $\neff$. When $\neff$ is increased the
Hubble parameter at recombination increases. In order to keep
the damping scale at the same value fixed by observations (see Eq.\ref{damp})
we need to decrease the free electron density at recombination. 
This can be achieved by simply accelerating the recombination process. 
This effect is clearly obtained by an increase in the fine structure
constant. This explains the direction of the degeneracy in contour plot.

When also a variation in the Helium abundance is considered,
the degeneracy changes direction. A larger value for $Y_p$
produces a large free electron fraction at recombination and
a smaller value for $\neff$ is needed to keep the damping scale
small. On the other hand a large value for $Y_p$ needs large 
values for $\alpha$. So now small values of $\neff$ are more compatible with
observations when $\alpha$ is larger.

\section{Conclusions}

In this paper we have presented new constraints on variation of the fine
structure constant from the latest CMB anisotropy measurements of the
ACT and SPT experiments, combined with other cosmological datasets.
We have found that assuming the standard value for $\neff$ and a
primordial Helium abundance of $Y_p=0.24$ the current data favours
a lower value for the fine structure constant at more than 
two standard deviations with $\alpha / \alpha_0 = 0.984 \pm 0.005$.

We have shown that this result relies on the assumption of the number
of relativistic degrees of freedom. When we let this parameter
vary freely, the standard value is again consistent with the
data considered. Varying also the primordial Helium content further
enlarges the error bars. Despite the existing degeneracies, the current
data offers the tantalizing suggestion of the presence of new
physics at the epoch of recombination.

Clearly, further experimental confirmation of the result is needed.
Fortunately, the results from the Planck satellite mission, expected
to be released early next year, will most probably clarify the issue.
The Planck experiment is indeed expected to have a
sensitivity of $\Delta \neff \sim 0.2$ and 
$\delta (\alpha / \alpha_0) \sim 0.002$ at $68 \%$ c.l. (see e.g. \cite{future}).

\section{Acknowledgments}
This work was supported by the PRIN-INAF grant 'Astronomy probes fundamental physics',
by the Italian Space Agency through the ASI contract Euclid- IC (I/031/10/0),
by FCT (Portugal) through grant PTDC/FIS/111725/2009 and by the
PHC-EGIDE/Programa PESSOA echange grant `Probing Fundamental Physics with Planck'
(FCT/1562/25/1/2012/S). The work of CJM is funded by a Ci\^ncia2007 Research
Contract, funded by FCT/MCTES (Portugal) and POPH/FSE (EC).

We acknowledge the hospitality of the University of the Azores during the
2011 Azores School on Observational Cosmology, at which the key ideas
behind this work were discussed.

\end{document}